\documentstyle[12pt]{article} \begin{document} 
\title{Fluctuations of the String Tension and Transverse Mass Distribution }

\author{A.Bialas \\ M.Smoluchowski Institute of Physics \\Jagellonian
University, Cracow\thanks{Address: Reymonta 4, 30-059 Krakow, Poland;
e-mail:bialas@thp1.if.uj.edu.pl}\\and \\Institute of Nuclear Physics, Cracow}

\maketitle

\begin{abstract} 
It is shown that Gaussian fluctuations of the string tension can account
for the  "thermal" distribution of transverse mass of particles
created in  decay of a colour string.
\end{abstract}

{\bf 1.} Recent precise data on the production rates of hadrons created in
$e^+e^-$ annihilation\footnote{A full list of data is given in
\cite{be,ch}.} allowed to analyse in detail the observed regularities.
The main general conclusion from these analyses is that 
production of hadrons can be very well explained by a (rather unexpected)
idea that they emerge from a thermodynamically equilibrated system.

This point was first recognized and emphasized some time ago by
Becattini \cite{be} who found that the particle spectra are consistent
with the model of two thermally equilibrated fireballs. The temperature
$T$ determined from the fit was found to be about 160 MeV.

Recently Chliapnikov \cite{ch} analysed again the experimental particle
distributions and found that they are also consistent with the thermal
model\footnote{I would like to thank B.Webber for calling my attention to 
this paper.}. In his picture hadrons emerge from a thermally
equilibrated system of (constituent) quarks and antiquarks. The fit to
the data gives $T \approx 140$ MeV.

It should be emphasized that all this is achieved with at most $3$
parameters, in contrast to the "standard" Monte Carlo codes like HERWIG
and JETSET which are much less effective in this respect\footnote{A
detailed discussion of this point can be found in \cite{ch}}.

These findings are difficult to reconcile with the 
generally accepted ideas
about hadron production in $e^+e^-$ collisions. The main difficulty is
that, since the process in question is expected to be rather fast, there
is hardly enough time for any equilibrium to set in. 

There is, however, a simple way to understand the findings of
refs. \cite{be,ch}:  The spectrum of primarily
produced partons (from which the final hadrons are formed) may be already so
close to the thermal one that there is no further need for secondary
collisions between partons to obtain the thermally equlibrated
distribution (both life-time of the system and the parton density are
irrelevant in this case). In the present note I shall argue that this
possibility may naturally occur in the string model.

{\bf 2.} In the string 
picture of hadron production in $e^+e^-$ annihilation \cite{ar,an} the
tranverse mass spectrum of the produced quarks (or diquarks) is taken 
from the Schwinger formula \cite{sch} which predicts \cite{br}
\begin{equation}
\frac {dn_{\kappa}}{d^2p_{\perp}} \sim e^{-\pi m_{\perp}^2/\kappa^2 }  \label{3}
\end{equation}
where $\kappa^2$ is the string tension
and $m_{\perp}$ is the transverse mass
\begin{equation}
m_{\perp}= \sqrt{p_{\perp}^2 +m^2}.   \label{2}
\end{equation}
On the other hand, the "thermal" distribution is exponential in $m_{\perp}$ 
\begin{equation}
\frac {dn}{d^2p_{\perp}} \sim e^{-m_{\perp}/T}   \label{1}
\end{equation}
rather than a Gaussian, as in (\ref{3}).

The main point of the present note is the observation  that Eq. (\ref{1}) can be 
reconciled with the Schwinger formula (\ref{3}) if the string tension 
 undergoes  fluctuations with the probability distribution in 
 the Gaussian form
\begin{equation}
P(\kappa) d\kappa =
\sqrt{ \frac{2}{\pi<\kappa^2>}} \exp\left(-\frac{\kappa^2}{2
<\kappa^2>}\right) d\kappa  \label{4}
\end{equation}
where $<\kappa^2>$ is the {\it average} string tension, i.e.
\begin{equation}
<\kappa^2> = \int_0^{\infty} P(\kappa) \kappa^2 d\kappa.   \label{5}
\end{equation}
Using (\ref{3}) and (\ref{4}) we thus have
\begin{equation}
\frac {dn}{d^2p_{\perp}} \sim
 \int_0^{\infty}d\kappa P(\kappa) e^{-\pi m_{\perp}^2/\kappa^2} =  
\frac{\sqrt{2}}{\sqrt{\pi<\kappa^2>}} \int_0^{\infty}d\kappa e^{-\frac{\kappa^2}{2 <\kappa^2>}} 
 e^{-\pi m_{\perp}^2/\kappa^2} \label{6}
\end{equation}
This integral can be evaluated using the identity \cite{abr}
\begin{equation}
 \int_0^{\infty}dt e^{-st} \frac{u}{2\sqrt{\pi t^3}}
e^{-\frac{u^2}{4t}}= e^{-u\sqrt{s}}.  \label{7}
\end{equation}
The result is
\begin{equation}
\frac {dn}{d^2p_{\perp}} \sim  \exp
\left(-m_{\perp}\sqrt{\frac{2\pi}{<\kappa^2>
}}\right)    \label{8}
\end{equation}
i.e. the "thermal" formula (\ref{1}) with
\begin{equation}
T= \sqrt{\frac{<\kappa^2>}{2\pi}}. \label{9}
\end{equation}

Using the standard value of the string tension $<\kappa^2>= 0.9$ Gev/fm, we
obtain $T=170$ MeV for the "temperature" of the primary partons, the 
value somewhat larger than those obtained by Beccatini \cite{be}
and Chliapnikov \cite{ch}. This is natural, as we expect that the
primary parton system may undergo some expansion (and thus cooling) 
before the final hadrons start to form.

We thus conclude that the possibility of a fluctuating string tension
may help to solve the apparent difficulty in the description of the mass
and transverse momentum spectra in the string model. The nature of the
fluctuations remains, however, an open question.

{\bf 3.} In search for a possible origin of such fluctuactions, it is
tempting to relate them to stochastic picture of the QCD vacuum studied
recently by the Heidelberg group \cite{do,na} \footnote{It was shown
that this picture helps to explain many features of the high-energy
cross-sections \cite{la}.}. In this approach the (average) 
string tension is given by the formula 
\begin{equation}
 <\kappa^2>= \frac {32 \pi k }{81} G_2 a^2   \label{10}
\end{equation}
 where $k$ is a constant ($k \approx .75$), $G_2$ is the
gluon condensate \cite{va} and $a$ is the correlation length of the
colour field in the vacuum (lattice calculations give $a=0.35$ fm
\cite{di}). This result has a natural
physical interpretation: It expresses the  string tension (i.e. energy per
unit lenght of the string) as a product of the vacuum energy density
(proportional to the gluon condensate) times the transverse area of the
string (proportional to $a^2$). 

In the stochastic vacuum model both quantities entering the R.H.S of
(\ref{10}) are expected to fluctuate. Indeed, the gluon condensate $G_2$
is proportional to the square of the field strength (its average value
can be estimated from studies of the charmonium spectrum \cite{va}).
Since the average value of the field strength in the vacuum must vanish,
it cannot be constant but changes randomly from point to point. Also
 $a^2$ represents only the average value of the fluctuating transverse
size of the string. Once this is accepted, it is natural to assume that
such fluctuations are described by a Gaussian distribution. This implies
fluctuations of the string tension in the form given by (\ref{4}).

An interesting consequence follows from this point of view.
First, the field fluctuations in the vacuum are expected to be
independent at the two space-time points whose distance exceeds the
correlation length $a$. This suggests that they may be also local {\it
along the string}\footnote{This was pointed out to me by W.Ochs.},
although the corresponding correlation length $a_s$ may be modified by
the presence of the $q\bar{q}$ pair which creates the string (thus $a_s$
may differ from $a$). This may have measurable effects. Indeed,
observation of a heavy $q\bar{q}$ or baryon-antibaryon pair at a point
along the string indicates that the string tension at this point took a
value well above the average. Thus by triggering on heavy particles one
may search for other effects of the large string tension, e.g.
increasing multiplicity in the neighbourhood. Since any effect of this
kind is limited to the region determined by the correlation length
 $a_s$, it may be possible to determine it experimentally and verify to
what extent it is different from the vacuum correlation length $a$ found
in lattice calculation \cite{di}. 

Needless to say, it would be very interesting to confirm or dismiss this
picture using the lattice QCD. This, however, does not seem to be an
easy task\footnote{I would like to thank J.Wosiek for  discussions about
this point.} .

Clearly, acceptance of the fluctuating string tension changes
many other features of the string model. It is, however, beyond the
scope of the present note to discuss them here.

{\bf 4.} In conclusion, we propose a modification of the original string
picture by introducing a fluctuating string tension. We have shown that
this assumption may help to explain the "thermal" nature of the spectra of
particles produced in $e^+e^-$ annihilation. We have also argued that it
seems justifiable in the stochastic picture of the QCD vacuum. It
remains an open and interesting question how this modification affects
the successful phenomenology of the string model.

\vspace{0.3cm}
{\bf Acknowledgements}
\vspace{0.3cm}

I am grateful to Wieslaw Czyz, Maciek Nowak, Wolfgang Ochs, Bryan Webber
and Jacek Wosiek for discussions which greatly helped to clarify the
idea presented in this note. This investigation was supported in part by
the KBN Grant No 2 P03B 086 14.

\end{document}